\documentclass{iau}
\usepackage{graphicx}
\usepackage{amsmath}
\usepackage{booktabs}
\usepackage{microtype}
\usepackage{wrapfig}

\defcitealias{PK25}{PK25}
\defcitealias{MZ20}{MZ20}
\defcitealias{PK26}{PK26}

\begin{document}

\lefttitle{Petr Kurf\" urst et al.}
\righttitle{What is the reason for the lack of cool evolved stars near the center of the Galaxy?}

\jnlPage{1}{7}
\jnlDoiYr{2026}
\doival{10.1017/xxxxx}

\aopheadtitle{Proceedings IAU Symposium}
\editors{M. Zaja\v{c}ek,  T. Je\v{r}\'{a}bkov\'{a}, V. Karas, R. Schödel \&  P. Sukov\'{a}, eds.}

\title{What is the reason for the lack of cool evolved stars near the center of the Galaxy?}

\author{Petr Kurf\"urst, Michal Zaja\v{c}ek, Norbert Werner, Ji\v{r}\'i Krti\v{c}ka, and Mat\'u\v{s} Labaj}
\affiliation{Department of Theoretical Physics and Astrophysics, Faculty of Science, Masaryk University, Kotl\'a\v{r}sk\'a 2, 611 37 Brno, Czech Republic}

\begin{abstract}
The observed deficit of bright late-type giants in the central parsec of the Milky Way remains an open problem. We investigate whether repeated passages of red giants (RGs) through a past jet from Sgr~A$^*$ can modify their envelopes and apparent spectral types. Three-dimensional hydrodynamical simulations follow a $1\,M_\odot$, $100\,R_\odot$ RG through up to ten jet crossings at $10^{-3}\,\text{pc}$, using jet kinetic luminosities of $10^{42}$, $10^{44}$, and $10^{48}\,\mathrm{erg\,s^{-1}}$. Each passage produces shocks, envelope ablation, and an asymmetric downstream tail. For jet luminosities up to $10^{44}\,\mathrm{erg\,s^{-1}}$, the cumulative ablated mass evolves approximately as $\Delta M\propto t^{1/2}$ and reaches about $10^{-4}\,M_\odot$ over a $10^5\,\text{yr}$  active phase. Repeated heating also raises the surface temperature from about $3600$ to $8500\,\text{K}$ during the first ten passages, potentially making an M-type giant appear as an A-type source. Jet feedback may therefore contribute to the apparent depletion of cool giants and the excess of hot stars in galactic nuclei (GN).
\end{abstract}
\maketitle

\begin{wrapfigure}{r}{0.33\textwidth}
\vspace{-0.4cm}
\includegraphics[width=0.33\textwidth]{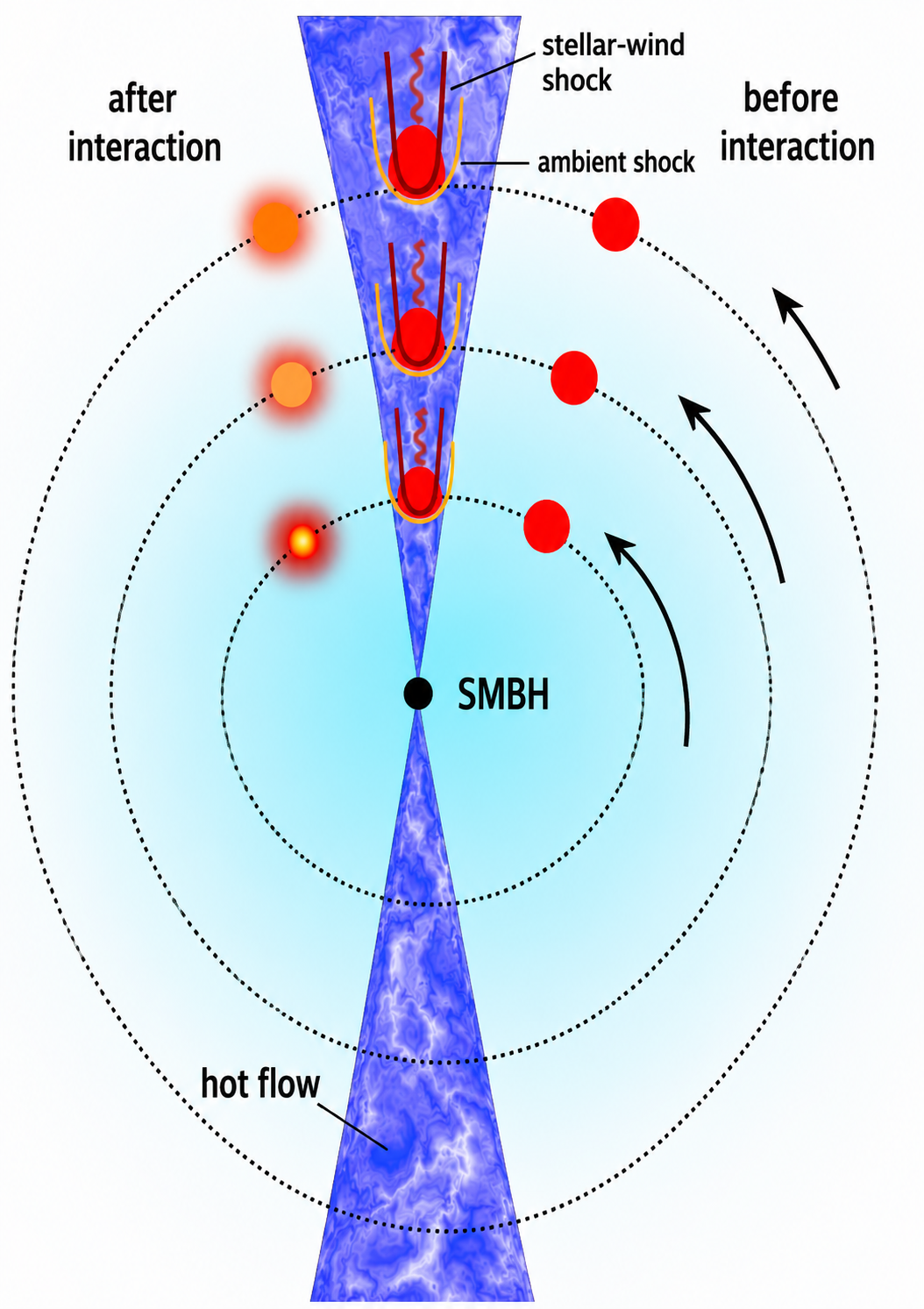}
\caption{Picture of an evolved star crossing a nuclear jet, producing an ambient bow shock and a second shock propagating into the stellar envelope. From \citetalias{MZ20} and \citetalias{PK25}.}
\label{fig:picture}
\end{wrapfigure}
\section{Motivation}
The central parsec of the Milky Way contains a dense nuclear star cluster (NSC) surrounding the $\simeq4\times10^6\,M_\odot$ supermassive black hole (SMBH) Sgr~A$^*$ \citep{Genzel2010,Schodel2020}. Although fainter late-type stars exhibit a cusp-like distribution, bright RGs and supergiants show a pronounced central depletion \citep{Schodel2020}. Proposed explanations include tidal stripping, star–disk collisions, star–star collisions, and interactions with a nuclear jet \citetext{\citealp{MZ20}, hereafter \citetalias{MZ20}}. These mechanisms need not be mutually exclusive; their relative importance is expected to vary with distance from the black hole and with the past activity of the GN.

A jet-induced mechanism is attractive because evolved stars have extended, weakly bound envelopes and slow winds. Their wind ram pressure is often insufficient to keep the jet stagnation point above the photosphere, so the jet can directly strike and ablate the stellar atmosphere \citetext{\citetalias{MZ20}; \citealp{PK25}, hereafter \citetalias{PK25}}. A star on an inclined orbit crosses both sides of the jet twice per orbital period. For an active phase of duration $t_{\rm jet}\sim10^5\,\text{yr}$, the expected number of crossings scales as
\begin{equation}
N_{\rm cross}\simeq \frac{t_{\rm jet}\sqrt{GM_\bullet}}{\pi r^{3/2}},
\end{equation}
which can reach thousands within the S-cluster region \citepalias{PK25}. The radius within which at least one encounter occurs is on the order of $2.6\,\text{pc}$ for the adopted Galactic center (GC) parameters \citepalias{PK25}. Repetition is therefore central to the problem: even modest changes during one passage can accumulate over many orbital cycles.

\section{Model and numerical method}
We model an RG of mass $M_\star=1\,M_\odot$ and initial radius $R_\star=100\,R_\odot$. A one-dimensional MESA model \citep{Paxton2011} is mapped to a polytropic envelope; the outer structure is best represented by a polytropic index near $n=3.68$. The dense core is represented by a softened central gravitational potential. Maintaining hydrostatic equilibrium for thousands of days is numerically demanding because spurious motions introduced by mapping a one-dimensional stellar model onto a Cartesian mesh can contaminate the weakly bound surface layers. To alleviate this problem, we follow the strategy of \citet{Ohlmann2017}. We employ a velocity-damping prescription and stabilize only the innermost region, $r<0.05R_\star$, with a smooth sponge term. The bulk of the envelope remains dynamically free and can respond to the jet.

The principal simulations are performed with CASTRO \citetext{\citealp{Almgren2010}; cf.~also~\citealp{PK26}, hereafter \citetalias{PK26}} in a cubic domain of side $2\,\text{au}$, initially discretized by $256^3$ cells and supplemented by AMR. The calculations solve the compressible hydrodynamic equations with self-gravity and an ideal-gas equation of state. The star is fixed at the domain center, while the orbital and jet motions are imposed through time-dependent inflow boundaries. The jet streams horizontally across the grid. Jet kinetic luminosities of $L_{\rm j}=10^{42}$, $10^{44}$, and $10^{48}\,\mathrm{erg\,s^{-1}}$ are explored, with velocities $v_{\rm j}=0.33c$ and $0.66c$. The main three-dimensional models place the star at $r=10^{-3}\,\text{pc}$ and follow up to ten crossings, corresponding to five orbits \citepalias{PK25}. A case at $10^{-2}\,\text{pc}$, for which the orbital period is much longer, is followed for two crossings in two dimensions.

Radiative cooling, thermal conduction, magnetic fields, and stellar rotation are not included. The resulting temperature evolution should consequently be interpreted as the adiabatic hydrodynamic response of the outer layers. These omissions are important for quantitative predictions but do not remove the central dynamical effects of repeated shock compression, stripping, and the exposure of deeper and hotter material.

\section{Hydrodynamical evolution}
When the RG enters the jet, the highly supersonic relative motion produces a bow shock in the jet material and a second shock driven into the stellar envelope \citetext{\citetalias{MZ20,PK25}; Fig.~\ref{fig:picture}}. The contact discontinuity between the shocked jet and shocked stellar gas wraps around the upstream hemisphere. Gas is diverted around the star and forms an elongated wake. Because the star also moves in its orbit, the downstream tail is not perfectly aligned with the jet axis but is deflected in the direction opposite to the orbital motion.

\begin{figure}[t]
\centering
\includegraphics[width=0.75\textwidth]{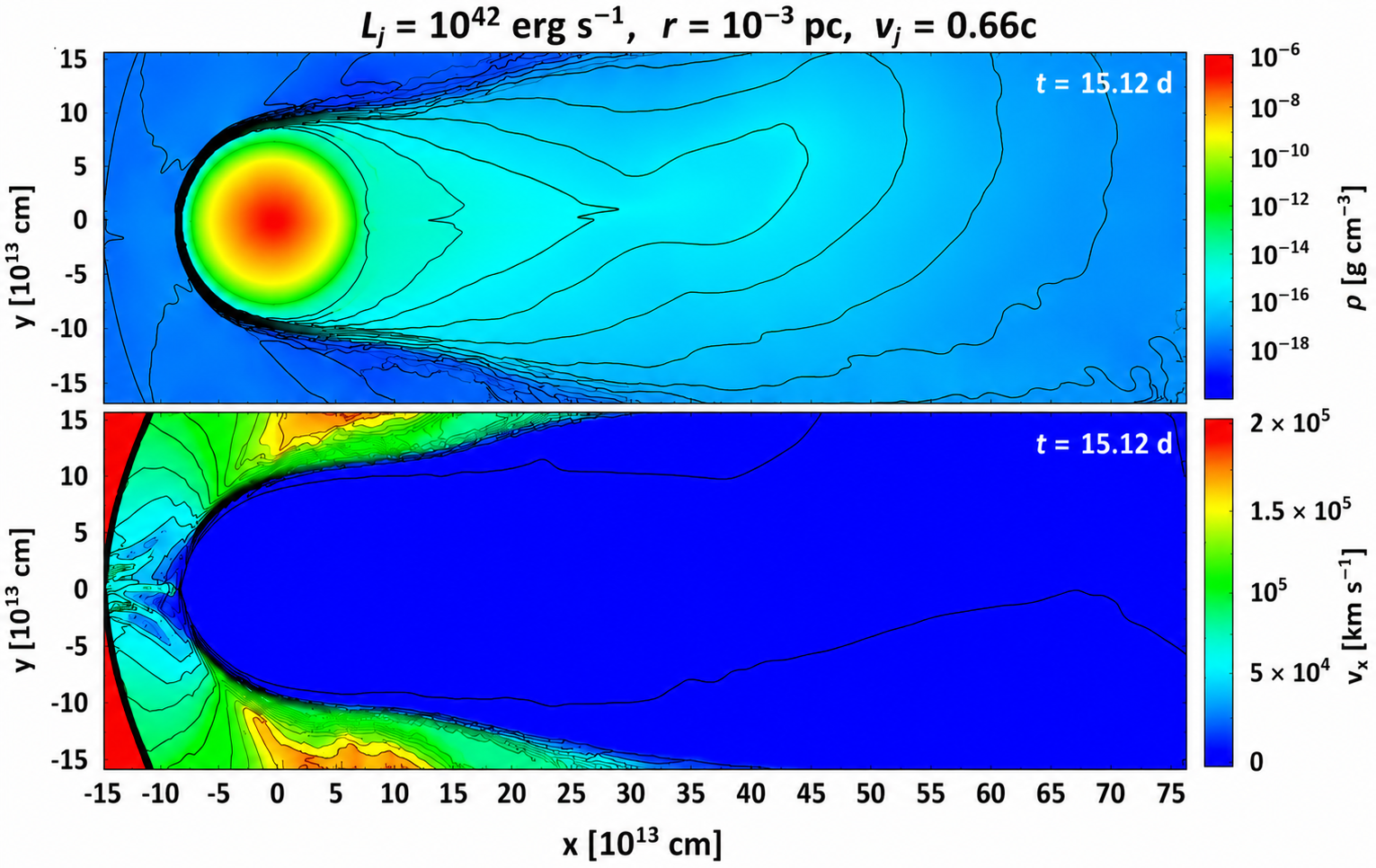}
\vspace{-0.2cm}
\caption{Density and longitudinal velocity maps show the mid-1st-crossing phase, position of the bow shock, and the asymmetric downstream tail for $L_{\rm j}=10^{42}\,\mathrm{erg\,s^{-1}}$, $r=10^{-3}\,\text{pc}$, and $v_{\rm j}=0.66c$. From \citetalias{PK25}.}
\label{fig:dynamics}
\end{figure}

The first entry rapidly disturbs the low-density surface layers. During the interval outside the jet, the star partly readjusts but does not return to its original state. At $10^{-3}\,\text{pc}$, the interval between consecutive crossings is only about $240\,\text{d}$, whereas the stellar Kelvin–Helmholtz (K-H) time is approximately $1.9\times10^2\,\text{yr}$ \citepalias{PK25}. The outer envelope, therefore, retains thermal and dynamical memory of earlier passages. By the tenth entry, the surface is more irregular, warmer, and more susceptible to further deformation.

The morphology is qualitatively similar over the considered luminosity range, but the amount of entrained jet material increases with jet density. At the highest luminosities, material temporarily accumulates inside the original stellar volume. This complicates the measurement of bound mass and can partially shield the stellar atmosphere, explaining why stronger jets do not necessarily produce proportionally larger ablation rates.

\section{Mass ablation}
The ablated mass is estimated by monitoring the change in mass within the volume initially occupied by the star. This volume-based diagnostic provides consistent comparisons among the simulations, although it does not perfectly separate bound stellar material from temporarily accumulated jet gas. For $L_{\rm j}=10^{42}\,\mathrm{erg\,s^{-1}}$ and $r=10^{-3}\,\text{pc}$, the fitted temporal behavior is close to \citepalias{PK25}
\begin{align}
\Delta M &\simeq A\left({t\over 1\,\mathrm{d}}\right)^{0.47-0.49},
&
A &\simeq3\times10^{-8}\,M_\odot.
\end{align}
The approximate square-root scaling persists for most models with $L_{\rm j}\leq10^{44}\,\mathrm{erg\,s^{-1}}$. Extrapolating the fitted trends gives a cumulative loss of order $10^{-6}\,M_\odot$ during a 10-year post-tidal disruption jet episode and $10^{-4}\,M_\odot$ during a $10^5$--year AGN phase \citepalias{PK25}. The dependence on whether $v_{\rm j}=0.33c$ or $0.66c$ is weak compared with other uncertainties.

\begin{figure}[t]
\centering
\includegraphics[width=0.87\textwidth]{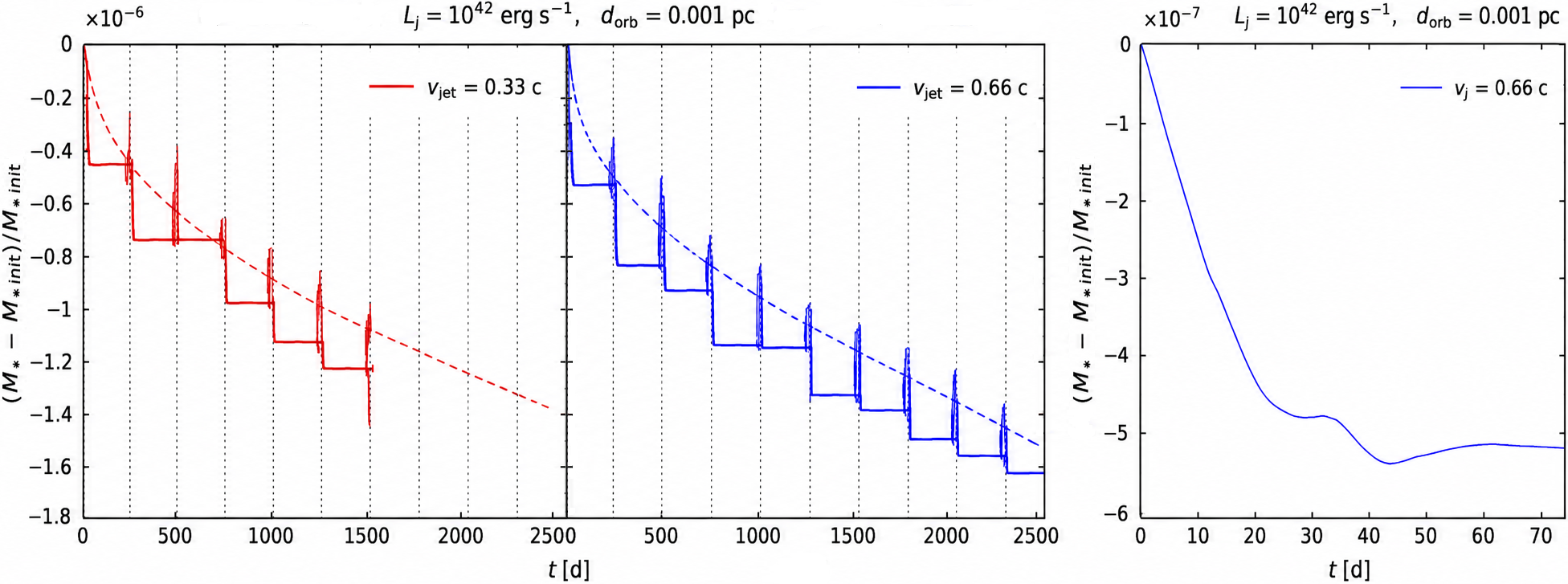}
\vspace{-0.2cm}
\caption{Relative mass loss over repeated star–jet crossings. For the relevant low-luminosity models, the long-term evolution follows approximately $\Delta M\propto \sqrt{t}$, with discrete, step-like increases associated with individual jet passages. The right panel shows a close-up of the first star–jet passage. From \citetalias{PK25}.}
\label{fig:massloss}
\end{figure}

The numerical and analytical estimates agree reasonably well for the lowest jet luminosity, but the simple analytical treatment of \citetalias{MZ20} strongly overpredicts ablation at larger $L_{\rm j}$ \citepalias{PK25}. Two effects contribute. First, dense jet gas can be temporarily packed around the star and counted within the original stellar volume. Secondly, the ablated surface becomes hotter, and its increased thermal pressure resists subsequent stripping. At $10^{48}\,\mathrm{erg\,s^{-1}}$, the fitted mass-loss evolution is markedly flatter, and the extrapolated loss over an AGN phase is only on the order of $10^{-5}\,M_\odot$. This non-monotonic behavior requires longer simulations and a diagnostic based directly on the gravitationally bound mass.

The total removed mass is too small to destroy an RG during one active episode, but it is comparable to the mass lost through the stellar wind over the same interval. Jet ablation can therefore enhance ordinary wind loss and, more importantly, continuously modify the thermodynamic state and observable appearance of the outer atmosphere.

\begin{wrapfigure}{R}{0.5\textwidth}
\centering
\includegraphics[width=0.5\textwidth]{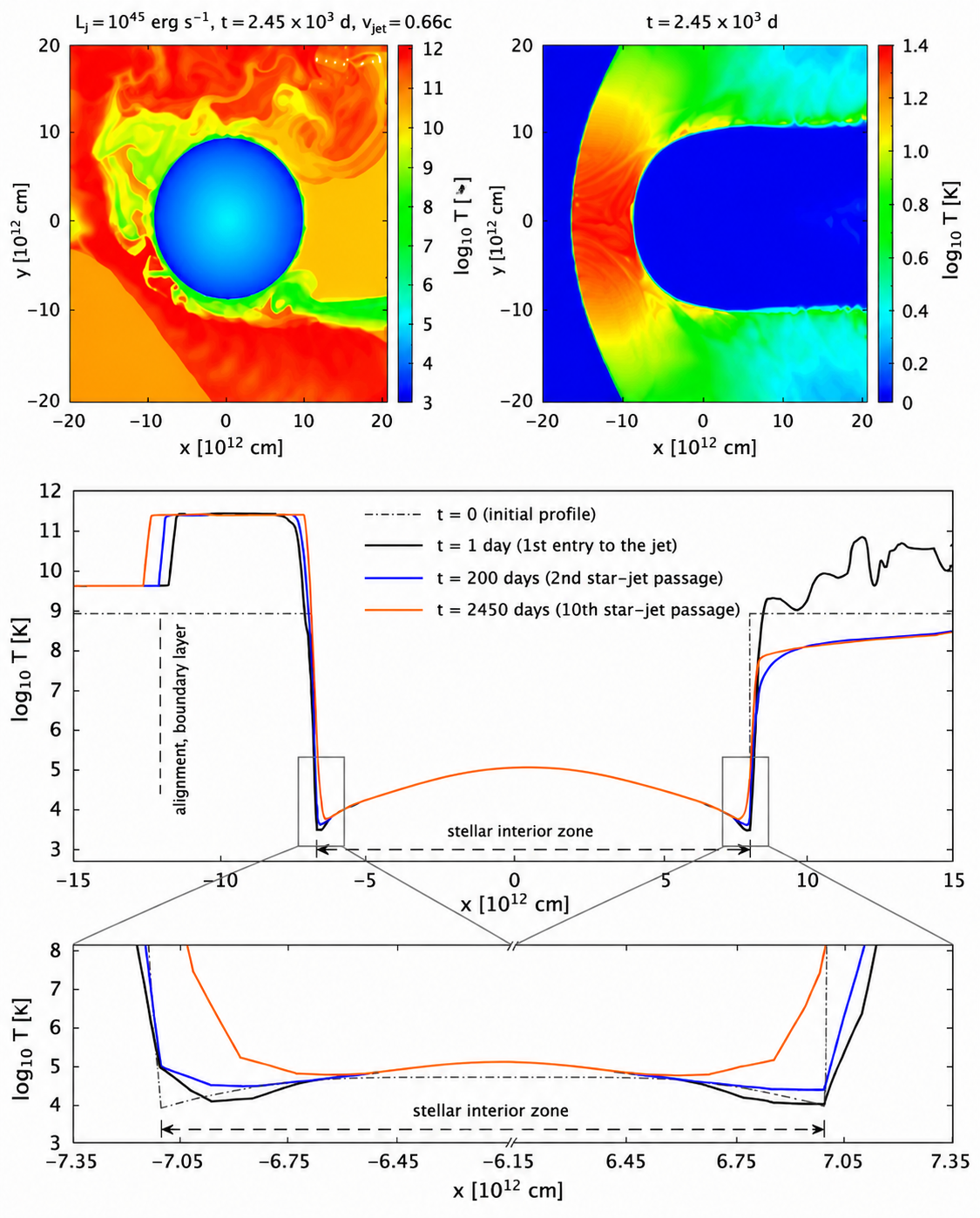}
\caption{Thermal evolution of the star: Repeated encounters expose and heat the outer layers, raising their characteristic temperature from about $3600$ to $8500\,\text{K}$ in the first ten passages. From \citetalias{PK25}.}
\label{fig:heating}
\end{wrapfigure}
\section{Surface heating and observable consequences}
The most striking result is the rapid heating of the stellar surface. In the $10^{-3}\,\text{pc}$ models, the characteristic temperature of the outer layers rises from approximately $3600\,\text{K}$ to about $8500\,\text{K}$ during the first ten crossings \citepalias{PK25}. The corresponding apparent spectral type may change from M to A; thus, an evolved star may cease to be identified observationally as a cool giant, even when only a small fraction of its total mass has been removed.

This mechanism may connect the deficit of bright cool giants with the apparent excess of hot stars, although tidal stripping, collisions, and disk interactions can operate concurrently. It also broadens the concept of AGN feedback: a nuclear jet can modify stellar envelope mass, temperature, luminosity, and apparent spectral type. After the jet switches off, the heated surface should cool on a K–H time scale of roughly 10–100 years. The affected fraction depends on orbital inclinations and jet geometry; preliminary calculations indicate stronger ablation for a half-opening angle of $5^\circ$ than for $10^\circ$, but longer simulations are required \citepalias{PK25}.

\section{Conclusions}
Repeated RG–jet encounters generate bow shocks, remove and heat the outer envelope, and form asymmetric wakes. For a plausible past Galactic jet, the cumulative loss grows approximately as $t^{1/2}$ and reaches $\sim10^{-4}\,M_\odot$ over $10^5$ yr. Although the removed mass is modest, ten passages can change an M-giant surface into an A-type-looking one. Jet-induced ablation may therefore contribute to the missing cool giants near Sgr~A$^*$ and leave observable records of past AGN activity. Longer calculations should determine the bound mass directly and include radiative cooling, magnetic fields, rotation, and a broader stellar population.

\acknowledgements
PK received support from the OPUS-LAP/GAČR-LA
bilateral project (2021/43/I/ST9/01352/OPUS22 and GF23-04053L). MZ and ML acknowledge the GA\v{C}R JUNIOR STAR grant no. GM24-10599M (``Stars in galactic nuclei: interrelation with massive black holes'') for support. Computational resources were supplied by the project 'e-Infrastruktura CZ' (e-INFRA LM2018140) provided within the programme Projects of Large Research, Development, and Innovation Infrastructures.
\begingroup
\renewcommand{\bibfont}{\fontsize{7.5}{8.5}\selectfont}

\endgroup


\begin{thebibliography}{99}
\bibitem[Almgren et al.(2010)]{Almgren2010} Almgren, A. S., et al. 2010, ApJ, 715, 1221
\bibitem[Genzel et al.(2010)]{Genzel2010} Genzel, R., Eisenhauer, F., \& Gillessen, S. 2010, Rev. Mod. Phys., 82, 3121
\bibitem[Kurf\"urst et al.(2025)]{PK25} Kurf\"urst, P., Zaja\v{c}ek, M., Werner, N., \& Krti\v{c}ka, J. 2025, MNRAS, 540, 1586
\bibitem[Kurf\"{u}rst et al.(2026)]{PK26} Kurf\"{u}rst, P., et al. 2026, A\&A, 710, A71
\bibitem[Ohlmann et al.(2017)]{Ohlmann2017} Ohlmann, S. T., et al. 2017, A\&A, 599, A5
\bibitem[Paxton et al.(2011)]{Paxton2011} Paxton, B., et al. 2011, ApJS, 192, 3
\bibitem[Sch\"odel et al.(2020)]{Schodel2020} Sch\"odel, R., et al. 2020, A\&A, 641, A102
\bibitem[Zaja\v{c}ek et al.(2020)]{MZ20} Zaja\v{c}ek, M., et al. 2020, ApJ, 896, 146
\end{thebibliography}
\end{document}